
\input phyzzx.tex
\def\a{\alpha}
\def\b{\beta}
\def\g{\gamma}

\def\d{\delta}
\def\D{\Delta}

\def\r{\rho}

\def\k{\kappa}
\def\l{\lambda}

\def\s{\sigma}

\def\pa{\partial}
\def\na{\nabla}
\def\hg{\hat g}

%

\def\pl#1{{\it Phys. Lett.} {\bf #1B}}

\def\prd#1{{\it Phys. Rev.} {\bf D#1}}

\def\np#1{{\it Nucl. Phys.} {\bf B#1}}

\def\mpl#1{{\it Mod. Phys. Lett.} {\bf A#1}}

%
\REF\call{C.G. Callan, S.B. Giddings, J.A. Harvey, and A. Strominger,
 \prd{45} (1992) R1005.}
\REF\dkd{F. David, \mpl{3} (1988) 1651, J. Distler and H. Kawai, \np{321}
(1989)
509.}
\REF\cham{A. Chamseddine,  \pl{256} (1991) 379, and \pl{258} (1991) 97,
 \np{368} (1992) 98; A Chamseddine and Th. Burwick,
preprint hepth 9204002. }
\REF\lenny{J.G. Russo, L.Susskind, and  L. Thorlacius, Stanford preprint (1992)
SU-ITP-92-4; L. Susskind and L. Thorlacius, Stanford preprint (1992)
SU-ITP-92-12. }
\REF\banks{T. Banks, A. Dabholkar, M.R. Douglas, and M O'Loughlin,  Rutgers
preprint RU-91-54.}
\REF\mtw{C.W. Misner, K. S. Thorne, and J.A. Wheeler,{\it Gravitation}  W. H.
Freeman  (1973).}
\REF\wit{E. Witten, \prd{44} (1991) 314.}
\REF\poly{A.M. Ployakov, \pl{103} (1981)207}
\REF\andy{A. Strominger, Santa Barbara preprint (1992),
UCSBTH-92-18,\hfill\break
hepth@xxx/9205028}

\pubnum {COLO-HEP-280\cr
 hepth@xxx/9205069}
\date={May, 1992}
\titlepage
\vglue .2in
\centerline{\bf Quantization of a Theory of 2d Dilaton Gravity}
\author{ S.P. de Alwis\foot{dealwis@gopika.colorado.edu}}
\address{Dept. of Physics, Box 390,\break
University of Colorado,\break Boulder, CO
80309}
\vglue .2in
\centerline{\caps ABSTRACT}

We discuss the quantization of the 2d gravity theory of Callan, Giddings,
Harvey, and Strominger (CGHS), following the procedure of David, and of
Distler and
Kawai. We find that the physics depends crucially on whether the number of
matter fields is greater than or less than 24. In the latter case the
singularity pointed out by several authors is absent but the
physical interpretation is unclear. In the former case (the one  studied by
CGHS) the quantum theory which gives CGHS in the linear dilaton semi-classical
limit, is different from that which gives CGHS in the extreme Liouville regime.

\endpage

Recently  Callan, Giddings, Harvey, and Strominger [\call ] (CGHS),  discussed
a model for two dimensional (dilaton) gravity coupled to matter. They showed
that classically the theory has solutions corresponding to collapsing matter
forming a black hole. This solution is in fact a linear dilaton flat metric
one, patched together with Witten's [\wit ] 2d black hole solution, along the
infall line of
a shock wave of 2d massless matter. In order to incorporate the quantum effects
(in lowest order) CGHS included the contribution of the conformal anomaly
coming from the conformally non-invariant measure in the matter sector path
integral.

In this paper we examine the consistency of this procedure. It is argued that
one way of carrying out the  quantization of the theory is to  follow the
procedure  of David, and of Distler and Kawai [\dkd ].\foot{Similar methods
have been
used in [\cham ]. However these works do not discuss the particular conclusions
for the
CGHS theory which is our main focus here. I wish to thank Dr. Chamseddine for
bringing
these references to my attention after an earlier version of this paper had
been circulated.}
Then we rediscover the
singularity pointed out in [\lenny , \banks ] when the
number $N$ of matter fields is greater than 24, and furthermore we find that
the quantum theory which leads to the CGHS action in the semi-classical linear
dilaton region is different from the the one which gives the CGHS action in the
extreme Liouville region. For $N<24$ there is no field space singularity but it
seems to lead to an unphysical theory with a negative flux of black hole
radiation.    The classical CGHS action\foot{we use MTW[\mtw] conventions} is

$$S={1\over 4\pi}\int d^2\s\sqrt{-g}[e^{-2\phi}(R+4(\na\phi )^2-4\l^2)-{1\over
2}\sum_{i=1}^N (\na f^i)^2 ].\eqn\cghs$$

where $\phi$ is the dilaton and $f^i$ are $N$ (unitary) matter
fields.\foot{This Lagrangian comes from the low energy limit of string theory.}
The corresponding quantum field theory is defined by

$$Z=\int {[dg]_g[d\phi ]_g[df]_g\over [Vol.~Diff.]} e^{iS[g,\phi,f]}\eqn\qft$$

The measures in the above path integral are derived from the metrics,
$$\eqalign{&||\d g||_g^2=\int d^2\s\sqrt{- g}g^{\a\g}g^{\b\d}(\d g_{\a\b}\d
g_{\g\d}
+\d g_{\a\g}\d g_{\b\d} )\cr
||\d\phi ||_g^2 &=\int d^2\s\sqrt{- g}\d\phi^2,\quad ||\d f||_g^2=\int
 d^2\sqrt{- g}\d_{ij}\d f^i\d f^j\cr}\eqn\met$$

 To evaluate the path integral one needs to gauge fix it. We choose the
conformal gauge $g=e^{2\r}\hat g$, where $\hg$ is a fiducial metric. Then the
path integral becomes,

 $$Z=\int([d\r ][d\phi ])_{\hg e^{2\r}}e^{iS(\phi,\r )}\D_f(e^{2\r}
 \hg )\D_{FP}(e^{2\r}\hg ),\eqn\fix$$

 where $S(\phi,\r )$ is the pure graviton-dilaton part of \cghs ,  the last
factor is the  Fadeev-Popov ghost determinant, and
  $$\D_f(e^{2\r }\hg )=\int [df]_{\hg e^{2\r}} e^{i S(f)},\eqn\mat$$

 $S(f)$ being the matter action.

 The measures in \fix , \mat , are again given by \met\ except that we must put
$g=e^{2\r}\hg$. In particular we have (up to a constant)

 $$||\d\r ||^2=\int d^2\s \sqrt{-\hg}\d\r^2.\eqn\metr$$

 From the well known transformation properties [\poly ] of the matter and ghost
determinants,

 $$\D_f(e^{2\r}\hg )\D_{FP}(e^{2\r}\hg )=\D_f(\hg )\D_{F.P.}(\hg )\exp
i[{N-26\over 6} S_L(\r,\hg ) +\mu\int d^2\s e^{2\r }\sqrt{-\hg}],\eqn\determ$$

 where
  $$S_L(\r,\hg )={1\over 4\pi }\int d^2\s\sqrt{-\hg}((\hat\na\r )^2+
 \hat R\r )\eqn\liouville$$.

  The quantum theory is then given by

  $$Z=\int ([d\r ][d\phi ])_{e^{2\r}\hg}[df]_{\hg}([db][dc])_{\hg}
  e^{iS(\r ,\phi ,f,\hg )+iS(b,c,\hg )}.\eqn\gfix$$
In the above equation $S(b,c,\hg )$ is the ghost action and

$$\eqalign{S(\r,\phi ,f,\hg )=&{1\over 4\pi}\int d^2\s\sqrt{-\hg}[e^{-2\phi}
(4(\hat\na\phi )^2-4\hat\na\phi .\hat\na\r )-\k\hat\na\r .\hat\na\r\cr
&-{1\over 2}\sum_1^N\hat\na f^i\hat\na f^i+\hat R(e^{-2\phi}-
\k\r )-4\l^2e^{2(\r-\phi )}]\cr}\eqn\cacn$$
 where $\k={26-N\over 6}$.
 For $\hg =\eta$ the Minkowski metric, this reduces to the CGHS action with
conformal anomaly term. \foot{equation (23) of [\call]  except that the ghost
contribution is ignored there.}

 There is however something strange about the path integral \gfix . The
  measures for matter and ghost are defined relative to the fiducial metric
$\hg $ while the $\r$ and $\phi$ measures are still defined in terms of the
original metric $g=e^{2\r}\hg$. In particular this means that the $\r$ measure
is not translationally invariant, and therefore that for example  the
(Dyson-Schwinger) quantum equation of motion gets modified from the equation
derived  from \cacn . In order to formulate the quantum theory in a manner
which yields a systematic semiclassical (or $1/N$) expansion it is necessary to
rewrite all measures in terms of the fiducial metric $\hg$. Thus we need to do
what David and Distler and Kawai [\dkd ] did for coformal field theory coupled
to $2d$ gravity.

 Assume (as in [\dkd ]) that the jacobian which arises in transforming to  the
 measures defined in terms of $\hg $ is of the form $e^{iJ}$ where $J$ is a
local renormalizable action in $\r$ and $\phi$. Putting $X^{\mu}=(\phi ,\r )$
we may write,

 $$Z=\int [dX^{\mu}]_{\hg}[df]_{\hg}([db][dc])_{\hg}e^{iI(X,\hg)+iS(f,\hg )+
 iS(b,c,\hg )}\eqn\part$$
  where
  $$I[X,\hg ]=-{1\over 4\pi}\int \sqrt{-\hg}[{1\over 2}
  \hg^{ab}G_{\mu\nu}\pa_aX^{\mu}\pa_bX^{\nu}+\hat R\Phi (X)
+T(X)].\eqn\sigmod$$

 In the above $G_{\mu\nu},\Phi$ and $T$ are functions of $X$ which are to be
determined and the measure $[dX^{\mu}]$ is derived from the natural metric on
the space $||\d X_{\mu}||^2=\int d^2\s \sqrt{-\hg}G_{\mu\nu}\d X^{\mu}\d
X^{\nu}$.

 The only a priori restriction on the functions $G,\Phi ,$ and $T,$ come from
the fact that $Z$ must be independent of the fiducial metric $\hg$, i.e. the
theory defined by the action $I+S_f+S_{b,c}$ with the standard translationally
invariant measures is a conformal field theory with zero central charge. So we
must satisfy the $\b$-function equations,
 $$\eqalign{\b_{\mu\nu}&={\cal R}_{\mu\nu} +2\na_{\mu}^G\pa_{\nu}\Phi-
 \pa_{\mu}T\pa_{\nu}T+\ldots\cr
 \b_{\Phi}&=-{\cal R}+4G^{\mu\nu}\pa_{\mu}\Phi\pa_{\nu}\Phi-4\na_G^2\Phi+
 {(N+2)-26\over 6}+G^{\mu\nu}\pa_{\mu}T\pa_{\nu}T-2T^2+\ldots\cr
 \b_T&=-2\na^2_GT+4G^{\mu\nu}\pa_{\mu}\Phi\pa{\nu}T-4T+\ldots\cr}\eqn\bet$$

 In the above ${\cal R}$ is the curvature of the metric $G$.
 These conditions are not sufficient to determine the functions uniquely, but
clearly they are necessary. If no further restrictions are imposed , they
define a class of quantum $2d$ dilaton-graviton theories. The analysis of CGHS
and others [\call ,\lenny ,\banks] will
 be valid provided that the functions $G,\Phi,$and $T,$ defined by \cacn\
satisfy \bet\ at least in the semiclassical regime.   Because of what happens
in the corresponding case studied in [\dkd ] we will make $\k$ (see \cacn ) a
parameter to be determined by
 \bet . Comparing \sigmod\ with \cacn\ we have,

 $$G_{\phi\phi}=-8e^{-2\phi},\quad G_{\phi\r}=4e^{-2\phi},\quad
G_{\r\r}=2\k,\eqn\metr$$
 $$\Phi =-e^{-2\phi}+\k\r ,\quad T=-4\l^2e^{2(\r-\phi)}.\eqn\dilt$$

 It is easy to see that the curvature ${\cal R}=0$. So we may transform to a
field space coordinate system which is Euclidean (or Minkowski). The
transformation
 $$ \r=\k^{-1}e^{-2\phi}+y\eqn\transf$$
 gives for the metric in field space

 $$\eqalign{ds^2&=-8e^{-2\phi}(d\phi^2-d\phi d\r )+2\k d\r^2\cr
               &=-{8\over\k}e^{-4\phi}(1+\k e^{2\phi})d\phi^2+2\k
dy^2\cr}\eqn\metric$$

In the latter  form we see (for $\k <0$) the singularity pointed out in
[\lenny, \banks ]. Now let us introduce a field space coordinate

$$ x=\int e^{-2\phi}(1+\k e^{2\phi })^{1\over 2}.\eqn\xcdt$$

Note that if $\k <0$ $x$ is real only in the "linear dilaton" region $\k
e^{2\phi}<1$. It is also convenient to introduce two more coordinates

$$X=2\sqrt{2\over|\k |}x,\quad Y=\sqrt{2|\k |}y.$$

Then we have,

$$ds^2=-{8\over\k}dx^2+2\k dy^2 = \mp dX^2\pm dY,\eqn\newmet$$

where the upper or lower signs are to be taken depending on whether $\k$ is
positive or negative respectively.\foot{ One may consider the coordinates $x,y$
(or $X,Y$) as the field space analog of Kruskal-Szekeres coordinates! Of course
the physical interpretation in terms of the original physical coordinates is
valid only outside the field space coordinate singularity.}
In the  Liouville region ${e^{-2\phi }\over
|\k |}$, we define the coordinate,
$$\bar X=2\sqrt 2\int dxe^{-x}\left (1+{e^{-2x}\over\k}\right )^{1\over
2}dx\eqn\trans$$
 we get
 $$ds^2=-d\bar X^2\pm dY^2\eqn\euclid$$
 As before the upper or lower signs are to be taken depending on whether $\k$
is positive or negative. Note that $X$ is real in the linear dilaton region and
imaginary in the Liouville region while the converse is true for $\bar X$.
 From \dilt and the above we also have the form of the dilaton in the new
coordinates,
 $$\Phi=\k y=\pm\sqrt{\pm \k\over 2}Y\eqn\lindil$$

 Thus in these new  coordinates we have a Euclidean (Minkowski) metric linear
dilaton theory in field space, and the first beta function equation \bet is
satisfied if we ignore quadratic terms in $T$. From the second equation in
\bet\ we then get

 $$\k={24-N\over 6}\eqn\kap$$

Thus the metric signature on field space as well as the absence or presence of
a singularity depends on whether
$N<24$ or $N>24$. As is well known the linear dilaton Euclidean (Minkowski)
metric theory is an exact solution of the  beta function equations , i.e. our
solution (with T=0) exactly satisfies the sufficiency criterion discussed in
the sentence before \bet .

Let us now discuss the tachyon $T$. We do not know how to incorporate the
contribution of the tachyon exactly. All we can do is to work to linear order
in $T$. Thus our discussion is valid only for $\l^2<< e^{2\phi}$.   By going to
the $x,y$ coordinate system \xcdt ,\newmet , we can solve the tachyon equation
exactly. In these coordinates we have,

 $${\k\over 4}\pa^2_xT-{1\over k}\pa_y^2T+2\pa_yT-4T=0.\eqn\taceq$$

  This has solutions of the form $e^{\b x+\a y}$, with
   ${\k\over 4}\b^2-{1\over k}\a^2+2\a-4=0$. Now we have to impose the boundary
condition that this solution goes over to the CGHS form given in \dilt\ in the
   semi-classical limit appropriate to the linear dilaton regime
$e^{-2\phi}>>|\k |$. Using the expansion of \xcdt\ and the expression for $y$
\transf , we find
$$T=-4\l^2e^{-{4\over\k}x+2y}=-4\l^2e^{2\r -2\phi}h(\k e^{2\phi}
),\eqn\tachyon$$

where $h(\k e^{2\phi})=1+O(\k e^{2\phi})$ and indeed can be written out exactly
{}.

  Now let us discuss the theory in the "Liouville region" $\k>e^{-2\phi}$.
The appropriate coordinates are $\bar X,Y$ defined in \trans . Solving the
tachyon equation in these coordinates and imposing the boundary condition that
the CGHS expression \dilt\ is reproduced in the extreme Liouville regime
$\k >>e^{-2\phi}>>1$ we find\foot{This is valid for $\k<0$. For $\k <0$ there
is a similar expression with $\cos\rightarrow\cosh$.}

$$T=-{\l^2\over 2}\k\left [\cos\left (X\right )e^{\sqrt{2\over |\k
|}Y}-e^{\a_+Y}\right ]\eqn\tacbar$$

where $\a_+==-{1\over 2}\sqrt{-2\k}+{1\over 2}\sqrt{-2\k +8},$
 By using the transformations \trans\ in the large $|\k |$ limit it is easily
seen that  $T$ goes over to the expression given in \dilt .  For $\k<0$, as in
the case studied by DKD the semi-classical expression for $T$ in the Liouville
region is
obtained
in the limit $N\rightarrow -\infty$.

The solution \tacbar\ is obviously quite different from the solution which goes
over to the CGHS value in the extreme linear dilaton region \tachyon .
What we have found is that  (for $\k <0$ ) we cannot have a quantum theory
which has the CGHS theory as its semi-classical limit in both the extreme
linear dilaton regime as well as the extreme Liouville region. This is already
obvious from the fact that the appropriate coordinates ($X,\bar X$) are real in
different regions (see discussion after \euclid ). The quantum theory (defined
with translationally invariant measures) which goes over to the CGHS theory in
the linear dilaton regime is given by,\foot{The theory  given by (27) is of the
Liouville type
and in fact can be solved. Also given that the Lioville theory is supposed to
be a conformal
theory it is likely that the same is true of (27) (i.e. to all orders in
$\l^2$). This solution and
its   physical implications are currently under investigation.}
$$Z=\int [dX][dY][df][db][dc]e^{iS[ X,Y,f]+iS_{ghost}},$$
 where,

 $$S={1\over 4\pi}\int d^2\s[\mp\pa_{+}X\pa_{-}
X\pm\pa_{+}Y\pa_{-}Y-\sum_i\pa_{+}f^i\pa_{-}f^i-T( X,Y)].\eqn\newaction$$

 with $T$ given by \tachyon\ whilst the theory which goes over to CGHS in the
extreme Liouville region is given by the above with $X$ replaced by $\bar X$
and $T$ given by \tacbar .

 What about the case $\k>0$. With the CGHS values for $G$ this would result in
a
 negative flux of Hawking radiation and therefore this solution is unphysical.
However it is possible that other solutions to $G$ exist which make this case
physical though the large $N$ analysis may remain problematic.
 Finally we note that the theory which goes over to CGHS one in the
semi-classical    linear dilaton region has  one wrong sign kinetic term
\newmet\ (for either sign of $\k$) but the theory has sufficient gauge
invariance (conformal invariance - Virasoro algebra) to gauge it away. On the
other hand in the theory (with $N>24$)  which gives CGHS in the large $N$ limit
we have two wrong sign fields and  the conformal symmetry is not sufficient to
gauge them both away.  However since neither the graviton nor the
dilaton are propagating modes,  the above probably does not mean that the
theory is non-unitary.

{\bf Note added} While  this work was being prepared for publication, a
preprint
by A. Strominger [\andy ] was received, in which the $N<24$ case  with what is
effectively a modified $G$, to avoid the problem of a negative flux of Hawking
radiation,  is discussed in some detail. This theory can in fact again be
written in the form
(27) with $T$ given by (25) except that the relation between $x,y$ and $\phi
,\r$ is
modified from (16) and (18) to

$\r=\k^1(e^{-2\phi}+4)+y,\quad dx=[(e^{-2\phi }-2)^2+\k (e^{-2\phi}-1)]^{1\over
2}d\phi$.

{\bf Acknowledgements}: The author is grateful to Leon Takhtajan for a
discussion.This work is partially
supported by Department of Energy contract No.
DE-FG02-91-ER-40672.

\refout
\end